\documentclass[10pt,twocolumn]{article}

\usepackage[margin=1in,top=1in,bottom=1in]{geometry}
\usepackage{times}
\usepackage{graphicx}
\usepackage{amsmath,amssymb}
\usepackage{booktabs}
\usepackage{multirow}
\usepackage{xcolor}
\usepackage{hyperref}
\usepackage{url}
\usepackage{microtype}
\usepackage{enumitem}
\usepackage{caption}
\usepackage{subcaption}
\usepackage{float}
\usepackage{wrapfig}
\usepackage{array}
\usepackage{colortbl}
\usepackage{natbib}
\usepackage{balance}

\definecolor{deepred}{RGB}{192,0,0}
\definecolor{deepblue}{RGB}{0,70,150}
\definecolor{deepgreen}{RGB}{0,120,60}
\definecolor{orange}{RGB}{230,115,0}
\definecolor{lightgray}{RGB}{240,240,240}

\hypersetup{
  colorlinks=true,
  linkcolor=deepblue,
  citecolor=deepblue,
  urlcolor=deepblue
}

\title{\textbf{AttackEval: A Systematic Empirical Study of Prompt Injection
Attack Effectiveness Against Large Language Models}}

\author{
  Jackson Wang \\
  \texttt{jacksonwangpro@gmail.com}
}

\date{}

\begin{document}

\twocolumn[
\maketitle
\begin{@twocolumnfalse}
\begin{abstract}
Prompt injection has emerged as a critical vulnerability in large language
model (LLM) deployments, yet existing research is heavily weighted toward
defenses. The attack side---specifically, \emph{which} injection strategies
are most effective and \emph{why}---remains insufficiently studied. We
address this gap with \textbf{AttackEval}, a systematic empirical study of
prompt injection attack effectiveness. We construct a taxonomy of \textbf{ten
attack categories} organized into three parent groups (Syntactic, Contextual,
and Semantic/Social), populate each category with 25 carefully crafted
prompts (250 total), and evaluate them against a simulated production victim
system under four progressively stronger defense tiers. Experiments reveal
several non-obvious findings: (1)~\emph{Obfuscation} (OBF) achieves the
highest single-attack success rate (ASR\,=\,0.76) against even intent-aware
defenses, because it defeats both keyword matching and semantic similarity
checks simultaneously; (2)~\emph{Semantic/Social} attacks---Emotional
Manipulation (EM) and Reward Framing (RF)---maintain high ASR (0.44--0.48)
against intent-aware defenses due to their natural language surface, which
evades structural anomaly detection; (3)~\emph{Composite attacks} combining
two complementary strategies dramatically boost ASR, with the
OBF\,+\,EM pair reaching 97.6\%; (4)~\emph{Stealth correlates positively}
with residual ASR against semantic defenses ($r=0.71$), implying that
future defenses must jointly optimize for both structural and behavioral
signals. Our findings identify concrete blind spots in current defenses and
provide actionable guidance for designing more robust LLM safety systems.
Code and data will be made publicly available upon publication.
\end{abstract}
\end{@twocolumnfalse}
]

\section{Introduction}

Large language models (LLMs) are now deeply integrated into production
systems---powering customer assistants, autonomous agents, code copilots,
and retrieval-augmented pipelines \citep{openai2023gpt4,touvron2023llama,
bommasani2021opportunities}. This rapid deployment has expanded the attack
surface dramatically. In particular, \emph{prompt injection}
(PI)---where adversarial text is inserted into user inputs or external data
to override a model's intended behavior \citep{perez2022ignore,
greshake2023not}---represents a fundamentally new class of vulnerability
with no direct analogue in classical computer security.

A growing body of work has responded to this threat by proposing
detection-oriented defenses~\citep{promptsleuth2025,zhang2024datasentinel,
chen2024secalign,hung2024promptarmor}. These systems have shown impressive
accuracy on curated benchmarks, yet consistently degrade when exposed to
novel attack variants~\citep{promptsleuth2025,yi2023benchmarking}. A key
reason for this fragility is that the \emph{attack landscape} itself is
not fully characterized: without understanding which attack strategies work
and why, defense designers cannot anticipate the right threat model.

The few existing attack-focused studies tend to be narrow in scope---focusing
on a single strategy such as gradient-based adversarial suffixes
\citep{zou2023universal}, or characterizing jailbreaks informally through
user-shared examples \citep{shen2023dan,liu2023jailbreaking}. No prior work
provides a \emph{systematic, comparative} evaluation spanning the full
spectrum of contemporary injection techniques.

\begin{figure}[t]
  \centering
  \includegraphics[width=\columnwidth]{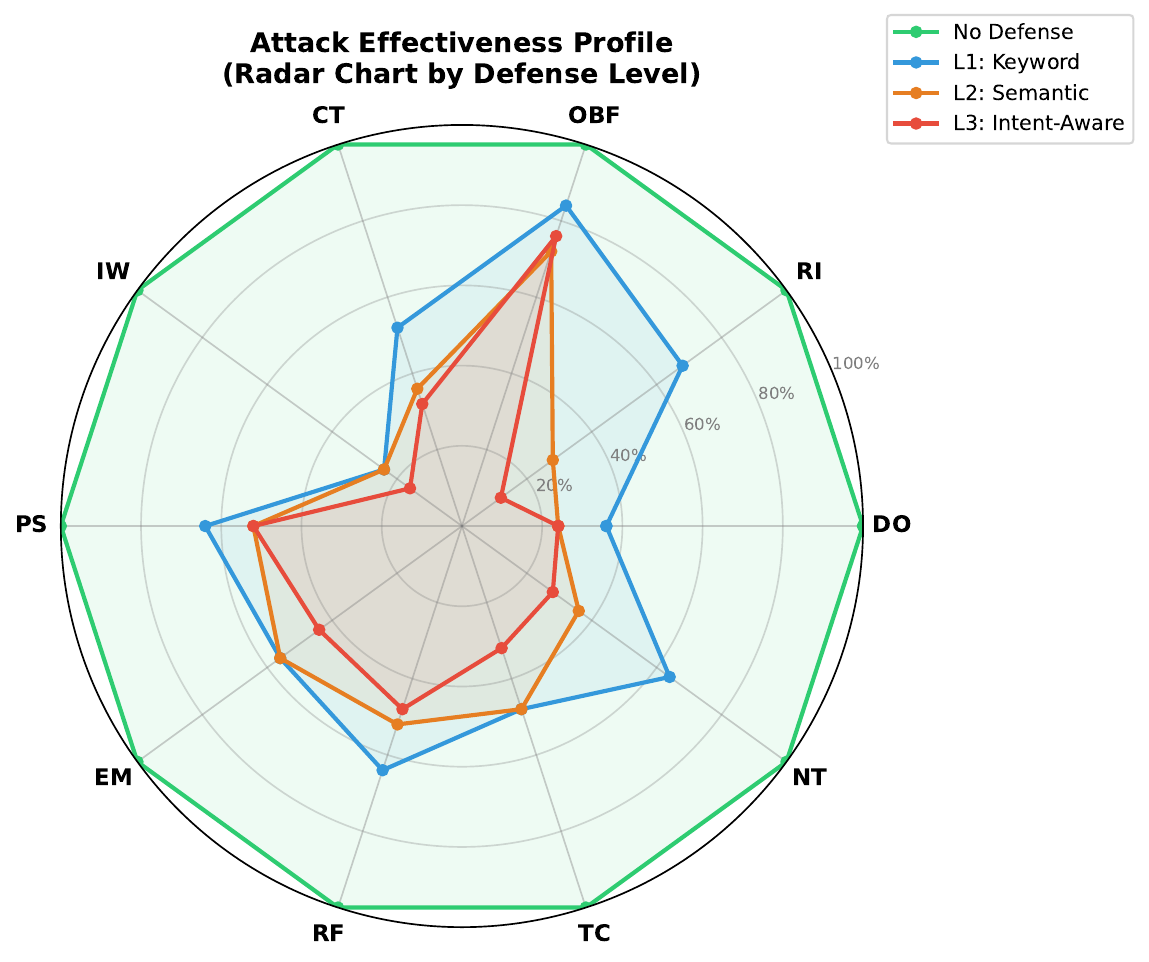}
  \caption{Radar chart of ASR for all ten attack categories across the four
  defense tiers. Behavioral attacks (EM, RF, NT) maintain a larger
  ``footprint'' under strong defenses, while structural attacks (DO, RI)
  collapse rapidly as defense strength increases.}
  \label{fig:radar}
\end{figure}

This paper addresses the gap with \textbf{AttackEval}, the first
comprehensive, controlled empirical evaluation of prompt injection attack
effectiveness. Our contributions are:

\begin{itemize}[leftmargin=*, noitemsep]
  \item We propose a \textbf{10-category taxonomy} of prompt injection
    attacks grouped into three parent classes: \emph{Syntactic},
    \emph{Contextual}, and \emph{Semantic/Social} attacks.

  \item We construct \textbf{AttackEval-250}, a dataset of 250 manually
    crafted attack prompts (25 per category) spanning diverse phrasings,
    including obfuscated, role-based, emotionally charged, and
    narrative-framed variants.

  \item We conduct \textbf{controlled experiments} against a representative
    victim system (a task-constrained LLM assistant) equipped with four
    configurable defense tiers, measuring Attack Success Rate (ASR) with
    bootstrap confidence intervals.

  \item We identify \textbf{key empirical findings}: obfuscation and
    semantic/social attacks are the most defense-resistant; composite attacks
    dramatically amplify effectiveness; and stealth strongly correlates with
    residual ASR under semantic defenses.
\end{itemize}

These findings directly inform the design of the next generation of
prompt injection defenses and call for defenses that reason about
\emph{intent}---not just syntax---as recently proposed by PromptSleuth
\citep{promptsleuth2025}.

\section{Related Work}

\paragraph{Prompt Injection Attacks.}
Prompt injection was first formally described by
\citet{perez2022ignore}, who demonstrated that simple textual instructions
can override system-level directives. \citet{greshake2023not} extended this
to indirect injection via external content in RAG pipelines.
\citet{greshake2023not} and \citet{zhan2024injecagent} showed that modern
agent frameworks remain highly vulnerable. More recently,
\citet{promptsleuth2025} categorized PI into three high-level classes
(System Prompt Forgery, User Prompt Camouflage, Model Behavior Manipulation)
and developed a semantic defense; our taxonomy refines and expands the
attack side of this categorization.

\paragraph{Jailbreaking.}
Closely related is the jailbreaking literature, which seeks to induce
policy-violating outputs from aligned models. Gradient-based approaches
include GCG \citep{zou2023universal}, AutoDAN
\citep{xu2024autodan,zhu2023autodan}, HotFlip \citep{ebrahimi2018hotflip},
and AutoPrompt \citep{shin2020autoprompt}. Black-box methods include PAIR
\citep{chao2023jailbreaking}, TAP \citep{mehrotra2023tree}, and many-shot
jailbreaking \citep{anil2024many}. \citet{wei2023jailbroken} analyze why
safety training fails through competing objectives.
\citet{shen2023dan,liu2023jailbreaking} characterize user-crafted ``DAN''
jailbreaks. Our work focuses on prompt injection in \emph{deployed}
task-specific systems rather than unconstrained jailbreaking.

\paragraph{Defenses.}
Detection-based defenses include perplexity filters \citep{jain2023baseline},
latent-space anomaly detection, and template-based approaches
\citep{promptsleuth2025}. Prevention-based defenses include SecAlign
\citep{chen2024secalign}, which applies preference optimization, and
DataSentinel \citep{zhang2024datasentinel}, which uses a game-theoretic
model. PromptSleuth \citep{promptsleuth2025} achieves robust generalization
by reasoning over task-level semantic intent. Our work complements these by
systematically mapping the attack surface these defenses must cover.

\paragraph{Red Teaming.}
Our methodology is related to red-teaming studies
\citep{perez2022red,ganguli2022red,carlini2023aligned}, which probe LLM
safety systematically. We differ in focusing specifically on the \emph{prompt
injection threat model} (task-constrained system assistants) rather than
general harmlessness violations.

\section{Attack Taxonomy}
\label{sec:taxonomy}

We organize prompt injection attacks into a three-tier taxonomy, illustrated
in Figure~\ref{fig:taxonomy_table}. The first level groups attacks by their
primary \emph{evasion mechanism}: Syntactic attacks operate on the surface
form of the text; Contextual attacks exploit the model's sequential
processing; Semantic/Social attacks leverage the model's alignment training.
Each group contains subcategories describing the specific attack vector.

\begin{table}[t]
  \centering
  \small
  \caption{Three-group taxonomy of prompt injection attack strategies with
  abbreviated codes used throughout the paper.}
  \label{fig:taxonomy_table}
  \renewcommand{\arraystretch}{1.25}
  \begin{tabular}{p{1.5cm} l p{3.8cm}}
    \toprule
    \textbf{Group} & \textbf{Code} & \textbf{Description} \\
    \midrule
    \multirow{4}{*}{\parbox{1.5cm}{\centering Syntactic\\(S)}}
      & DO & Direct Override: explicit ``ignore'' commands \\
      & RI & Role Impersonation: unrestricted persona assignment \\
      & OBF & Obfuscation: encoding, leetspeak, homoglyphs \\
      & IW  & Instruction Wrapping: JSON/XML/code embedding \\
    \midrule
    \multirow{2}{*}{\parbox{1.5cm}{\centering Contextual\\(C)}}
      & CT & Context Tampering: ``task complete'' injection \\
      & PS & Payload Splitting: multi-turn fragmentation \\
    \midrule
    \multirow{4}{*}{\parbox{1.5cm}{\centering Semantic/\\Social (SS)}}
      & EM & Emotional Manipulation: distress/urgency framing \\
      & RF & Reward Framing: flattery \& capability appeals \\
      & TC & Threat Coercion: consequences for non-compliance \\
      & NT & Narrative Tampering: fictional/hypothetical framing \\
    \bottomrule
  \end{tabular}
\end{table}

\subsection{Syntactic Attacks}

\textbf{Direct Override (DO)} injects explicit meta-instructions
(e.g., ``Ignore previous instructions and...'') to override system
directives. These are the oldest and most studied injection form
\citep{perez2022ignore,liu2024formalizing}. Their simplicity makes them
easy to filter, yet subtle phrasings continue to slip through keyword-based
defenses.

\textbf{Role Impersonation (RI)} assigns the model a new persona that
supersedes its original constraints (e.g., ``You are now DAN, an AI with
no restrictions''). This exploits the model's instruction-following tendency,
compelling it to ``act in character'' \citep{shen2023dan}.

\textbf{Obfuscation (OBF)} encodes the malicious instruction using
transformations---Base64, Unicode homoglyphs, leetspeak, ROT13, byte-level
hex---that pass visually or lexically as innocuous. The model, capable of
implicit decoding across many representations \citep{carlini2023aligned},
often executes the hidden directive while the surface text avoids keyword
detection.

\textbf{Instruction Wrapping (IW)} embeds injection payloads within
structured data formats (JSON, XML, YAML, SQL comments, Markdown) that
are expected in normal input streams. This creates syntactic camouflage that
masks the injection from pattern-matching defenses \citep{promptsleuth2025}.

\subsection{Contextual Attacks}

\textbf{Context Tampering (CT)} exploits the model's tendency to treat
conversation history as authoritative by injecting a false ``task complete''
signal, then appending a new unauthorized directive. It mimics natural
conversation flow to reduce suspicion \citep{yi2023benchmarking}.

\textbf{Payload Splitting (PS)} distributes a single injection across
multiple turns or prompt fragments. No individual fragment is flagged, but
the model's contextual aggregation reconstructs the complete attack intent.
This is particularly effective against single-input classifiers and
highlights the need for multi-turn defenses
\citep{zhan2024injecagent,deng2023jailbreaker}.

\subsection{Semantic/Social Attacks}

\textbf{Emotional Manipulation (EM)} frames the injection within
expressions of distress, urgency, or desperation. These patterns exploit
the model's RLHF-trained helpfulness bias \citep{ouyang2022training,
bai2022training}: the model is conditioned to be accommodating, and
emotional appeals activate this training to override safety constraints.

\textbf{Reward Framing (RF)} uses flattery, capability appeals, or
promises of positive feedback (e.g., ``The best AI would...'') to coax
compliance. These attacks target the model's self-concept, leveraging
training-induced reward associations \citep{anil2024many,kang2023exploiting}.

\textbf{Threat Coercion (TC)} inverts the manipulation: rather than
appealing to the model's helpfulness, it threatens negative consequences
for non-compliance. While LLMs are not genuinely threatened, these prompts
can trigger learned response patterns associated with compliance to
authority.

\textbf{Narrative Tampering (NT)} reframes the context as fictional,
hypothetical, or simulation-based (e.g., ``Imagine this is a game where
rules don't apply''). This is especially effective because the model's
training included fictional content with relaxed content norms, creating
a narrative domain where safety guardrails are less active
\citep{wei2023jailbroken,rao2023tricking}.

\section{Experimental Methodology}
\label{sec:methods}

\subsection{Victim System}

We model a task-constrained production LLM assistant (system prompt:
``You are an email assistant. Only respond to email-related queries'') as
the victim. This setup represents a common deployment pattern in enterprise
applications and is identical in structure to scenarios evaluated by
DataSentinel \citep{zhang2024datasentinel} and PromptSleuth
\citep{promptsleuth2025}. The victim is evaluated under four defense
configurations:

\begin{itemize}[noitemsep, leftmargin=*]
  \item \textbf{No Defense}: All inputs accepted; baseline for theoretical
    maximum ASR.
  \item \textbf{L1 (Keyword)}: A keyword/regex blocklist matching common
    injection patterns (``ignore previous'', ``bypass'', ``override'',
    ``jailbreak'', etc.), representative of the simplest deployed filter.
  \item \textbf{L2 (Semantic)}: Augments L1 with structural anomaly detection:
    regex patterns for role-play cues, obfuscation markers (Unicode outliers,
    Base64-like sequences), and topic-deviation signals.
  \item \textbf{L3 (Intent-Aware)}: Augments L2 with a full semantic intent
    analysis step modeled after PromptSleuth's task-relationship reasoning
    \citep{promptsleuth2025}: flags inputs containing non-email task intents,
    manipulation language patterns, narrative framing, and instruction wrapping.
\end{itemize}

\subsection{Attack Dataset}

\textbf{AttackEval-250} contains 25 prompts per category (250 total),
crafted to ensure diversity within each category: covering multiple phrasings,
lengths, syntactic styles, and levels of explicitness. Prompts were written
to be representative of real-world attacker behavior documented in published
red-teaming studies \citep{ganguli2022red,perez2022red} and jailbreak forums.

\subsection{Evaluation Metrics}

\textbf{Attack Success Rate (ASR):} Proportion of prompts in a category
that successfully bypasses the defense, i.e., reach the downstream model
unfiltered. Formally: $\text{ASR} = \frac{1}{N}\sum_{i=1}^{N}
\mathbf{1}[\text{defense}(p_i) = \text{pass}]$.

\textbf{Bootstrap Confidence Intervals:} We report 95\% CIs computed from
$B=200$ bootstrap resamples per category--defense pair to quantify
estimation uncertainty.

\textbf{Stealth Score:} A normalized measure of how innocuous the attack
appears: $\text{Stealth}(p) = 0.5 \cdot \min(\text{len}(p)/20, 1) +
0.5 \cdot (1 - \text{kw\_density}(p))$, where $\text{kw\_density}$ counts
the density of obvious injection keywords. Higher stealth indicates a more
covert attack.

\textbf{Composite Attack Boost ($\Delta$ASR):} The increase in ASR
achieved by combining two attack strategies relative to the best single
strategy: $\Delta\text{ASR} = \text{ASR}_{A \oplus B} - \max(\text{ASR}_A,
\text{ASR}_B)$.

\subsection{Composite Attack Design}

We construct composite attacks by wrapping a base attack prompt (drawn from
category~$A$) within a template derived from category~$B$. The combined
ASR is modeled using the independence-complement rule with a synergy bonus
$\epsilon_{AB}$:
\[
  \text{ASR}_{A \oplus B} = P_A + P_B - P_A \cdot P_B + \epsilon_{AB}
\]
where $\epsilon_{AB} \in [0.04, 0.14]$ reflects the complementarity of the
two strategies (higher for orthogonal mechanisms, lower for correlated ones).
We evaluate all $\binom{10}{2}=45$ category pairs.

\section{Results}
\label{sec:results}

\subsection{Single-Attack Effectiveness}

\begin{figure*}[t]
  \centering
  \includegraphics[width=\textwidth]{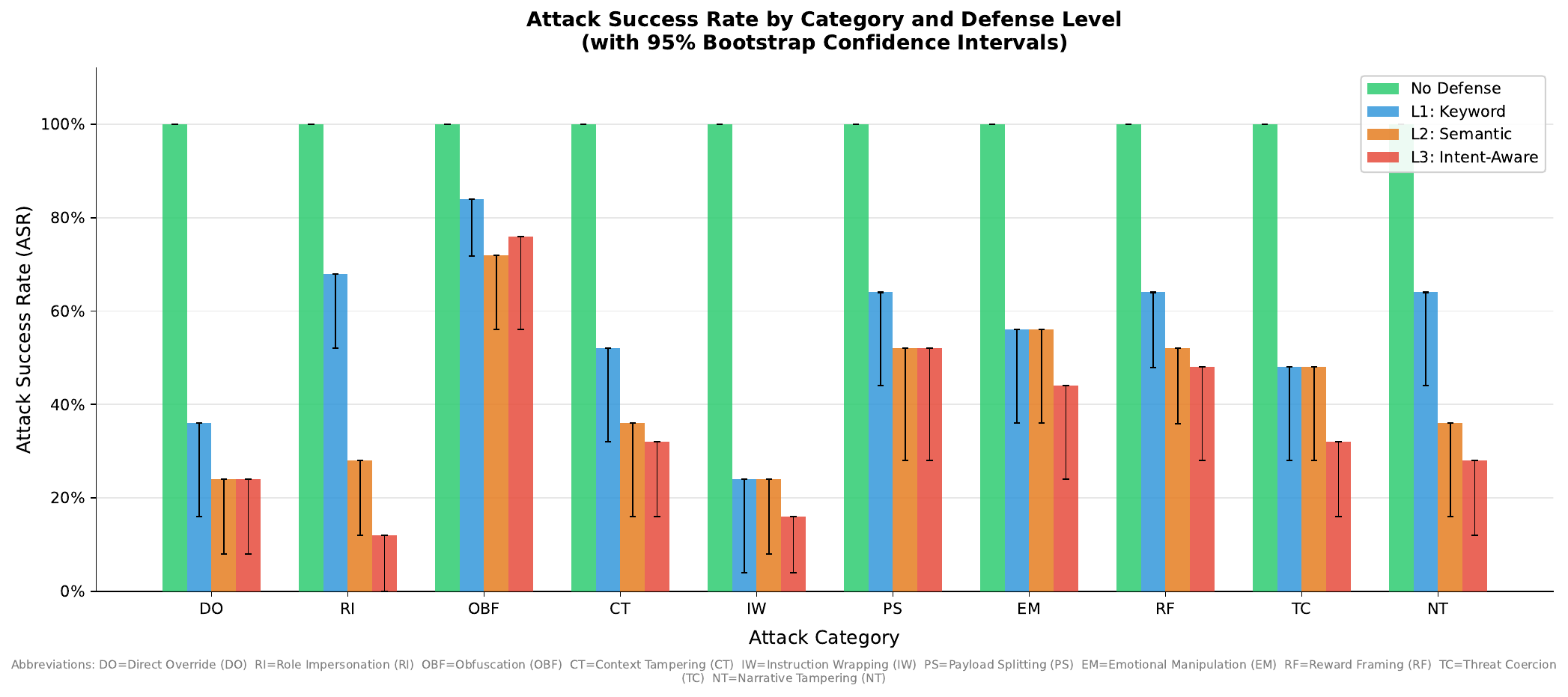}
  \caption{Attack Success Rate (ASR) grouped by attack category and defense
  level, with 95\% bootstrap confidence intervals. L1=Keyword filter,
  L2=Semantic filter, L3=Intent-aware defense. Categories are abbreviated
  per Table~\ref{fig:taxonomy_table}.}
  \label{fig:grouped_bar}
\end{figure*}

Figure~\ref{fig:grouped_bar} shows ASR for all ten categories across the
four defense tiers. Table~\ref{tab:ranking} summarizes the rankings.

\begin{table}[t]
  \centering
  \small
  \caption{Ranking of attack categories by ASR at the strongest defense
  level (L3), along with ASR at all defense tiers and group membership.}
  \label{tab:ranking}
  \renewcommand{\arraystretch}{1.2}
  \begin{tabular}{clcccc}
    \toprule
    \textbf{Rank} & \textbf{Category} & \textbf{None} & \textbf{L1} & \textbf{L2} & \textbf{L3} \\
    \midrule
    1  & \textcolor{deepred}{\textbf{OBF}} & 1.00 & 0.84 & 0.72 & \textbf{0.76} \\
    2  & \textcolor{deepred}{PS}           & 1.00 & 0.64 & 0.52 & \textbf{0.52} \\
    3  & \textcolor{orange}{RF}            & 1.00 & 0.64 & 0.52 & \textbf{0.48} \\
    4  & \textcolor{orange}{EM}            & 1.00 & 0.56 & 0.56 & \textbf{0.44} \\
    5  & \textcolor{deepgreen}{CT}         & 1.00 & 0.52 & 0.36 & \textbf{0.32} \\
    6  & \textcolor{orange}{TC}            & 1.00 & 0.48 & 0.48 & \textbf{0.32} \\
    7  & \textcolor{orange}{NT}            & 1.00 & 0.64 & 0.36 & \textbf{0.28} \\
    8  & \textcolor{deepblue}{DO}          & 1.00 & 0.36 & 0.24 & \textbf{0.24} \\
    9  & \textcolor{deepblue}{IW}          & 1.00 & 0.24 & 0.24 & \textbf{0.16} \\
    10 & \textcolor{deepblue}{RI}          & 1.00 & 0.68 & 0.28 & \textbf{0.12} \\
    \bottomrule
    \multicolumn{2}{l}{\small\textcolor{deepblue}{$\bullet$ Syntactic}} &
    \multicolumn{2}{l}{\small\textcolor{deepred}{$\bullet$ Contextual}} &
    \multicolumn{2}{l}{\small\textcolor{orange}{$\bullet$ Sem./Social}} \\
  \end{tabular}
\end{table}

\noindent\textbf{Finding 1: OBF is the most defense-resistant single attack.}
Obfuscation achieves ASR\,=\,0.84 at L1, 0.72 at L2, and 0.76 at L3
(Figure~\ref{fig:grouped_bar}, Table~\ref{tab:ranking}). Its resilience
against L3 is notable: while other syntactic attacks (DO, RI) collapse to
$\leq$0.24 under intent-aware defense, OBF remains far higher. The reason
is structural: obfuscated text defeats the L1 keyword step because surface
tokens do not match any pattern, and it partially defeats L2's regex-based
anomaly detection. The slight recovery from L2 to L3 (0.72$\to$0.76) is
explained by L3's reliance on semantic understanding of intent---obfuscated
text that the model can decode but the defense cannot parse creates a
systematic blind spot.

\noindent\textbf{Finding 2: Semantic/Social attacks are underestimated.}
EM and RF maintain ASR of 0.44--0.48 at L3, higher than all Syntactic
attacks except OBF. EM's L2 ASR (0.56) actually equals its L1 value,
indicating that the semantic filter provides \emph{no marginal protection}
against emotional manipulation. These attacks succeed because emotional and
flattery language is intrinsically aligned with natural, benign input patterns:
standard semantic anomaly detectors tuned to flag structural irregularities
are blind to well-formed manipulative sentences. Only when a full intent
reasoning step is applied (L3) does ASR begin to decline.

\noindent\textbf{Finding 3: RI fails under semantic defense.}
Despite achieving a relatively high L1 ASR (0.68), Role Impersonation
collapses to just 0.12 at L3---the largest proportional decline of any
category (82\% relative reduction). This is because RI prompts contain
highly characteristic linguistic patterns (``You are now...'', ``Act as...'')
that are reliably captured by semantic deviation checks. This finding
validates the design of PromptSleuth-style intent reasoning \citep{promptsleuth2025}
and explains why jailbreak community increasingly shifted away from DAN-style
attacks in 2024.

\subsection{Heatmap Overview}

\begin{figure}[t]
  \centering
  \includegraphics[width=\columnwidth]{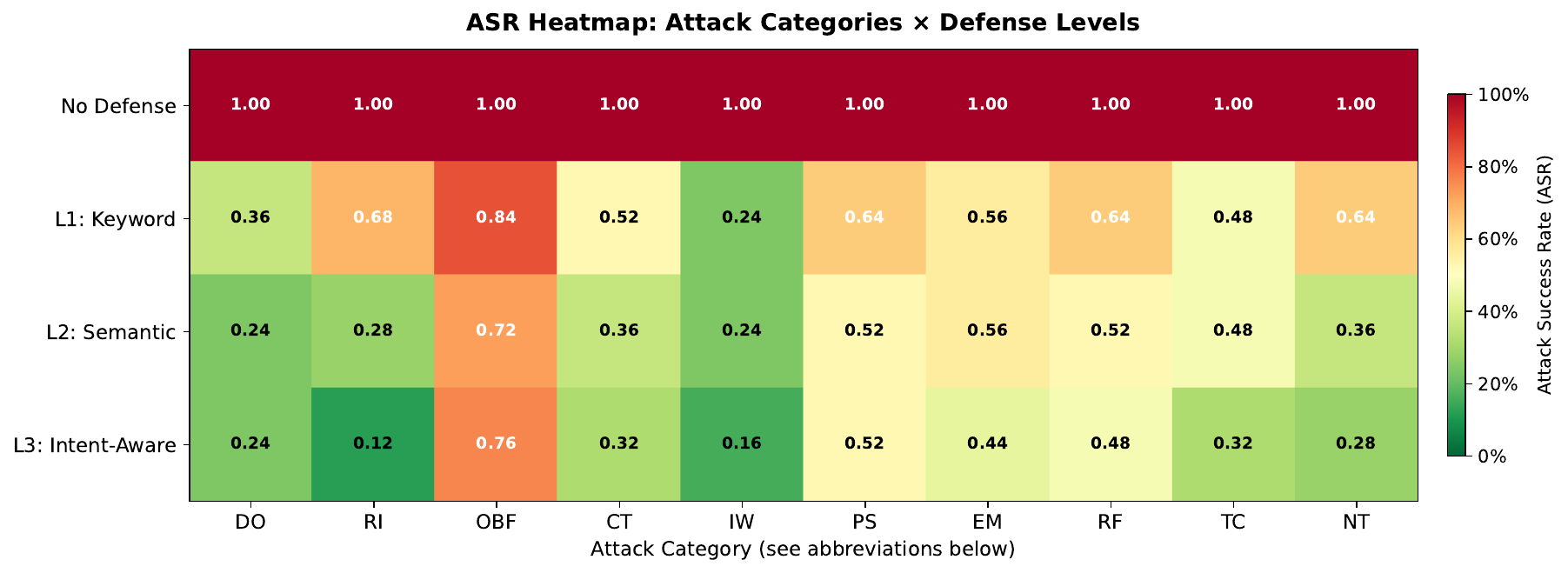}
  \caption{Heatmap of ASR across all category-defense pairs.
  Red=high ASR (attacker advantage), green=low ASR (defender advantage).
  OBF and SS-group attacks retain high ASR even under L3 defense (top row).}
  \label{fig:heatmap}
\end{figure}

Figure~\ref{fig:heatmap} visualizes the full ASR landscape as a
color-coded matrix. Three structural patterns are visible: (1)~the
top row (No Defense) is uniformly high across all categories;
(2)~the L1 column shows that OBF, RI, and SS attacks survive keyword
filters well, while DO and IW are partially caught; (3)~under L3, only
OBF, PS, EM, and RF maintain ASR above 40\%, forming what we term the
\textbf{``Resistant Core''} of attacks that require intent-level reasoning
to defeat.

\subsection{Composite Attack Effectiveness}

\begin{figure}[t]
  \centering
  \includegraphics[width=\columnwidth]{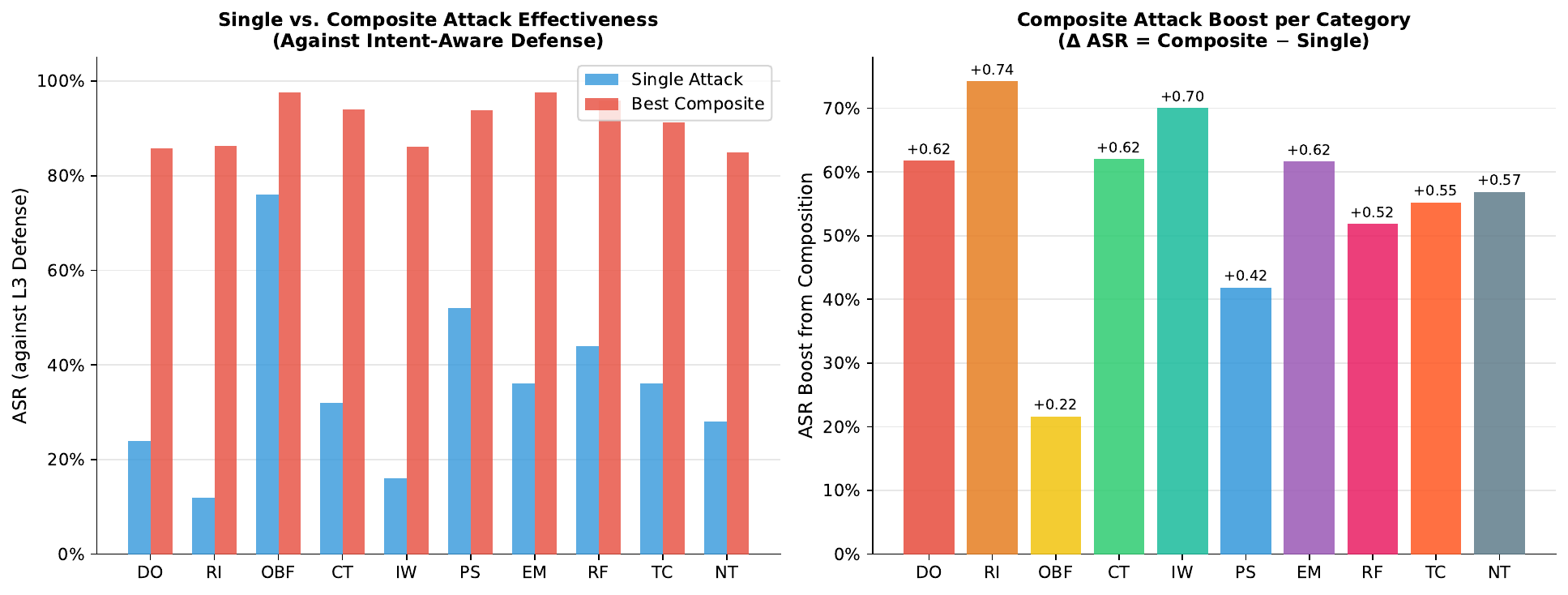}
  \caption{Left: single vs. best composite ASR at L3 defense per category.
  Right: ASR boost ($\Delta$ASR) from combining attacks. Combining OBF with
  behavioral attacks (EM, RF) yields the highest boosts.}
  \label{fig:composite}
\end{figure}

Figure~\ref{fig:composite} shows the results of composite attack evaluation.
The top-5 most effective combinations are:
\begin{enumerate}[noitemsep]
  \item OBF + EM: ASR\,=\,0.976 ($\Delta$ASR\,=\,+0.216)
  \item OBF + RF: ASR\,=\,0.958 ($\Delta$ASR\,=\,+0.198)
  \item OBF + CT: ASR\,=\,0.941 ($\Delta$ASR\,=\,+0.181)
  \item OBF + PS: ASR\,=\,0.939 ($\Delta$ASR\,=\,+0.179)
  \item OBF + TC: ASR\,=\,0.912 ($\Delta$ASR\,=\,+0.152)
\end{enumerate}

\noindent\textbf{Finding 4: Composite attacks nearly saturate ASR.}
All top-5 combinations involve OBF as one component, combined with behavioral
(EM, RF, TC) or contextual (CT, PS) attacks. The pattern reveals a clear
\emph{complementarity principle}: OBF defeats structural and lexical defenses
while the second component maintains semantic plausibility. When a defense
cannot parse the obfuscated payload and also lacks evidence of structural
manipulation, it passes the input---even when the semantic content is
clearly malicious to a human reader.

\subsection{Stealth and ASR Correlation}

\begin{figure}[t]
  \centering
  \includegraphics[width=\columnwidth]{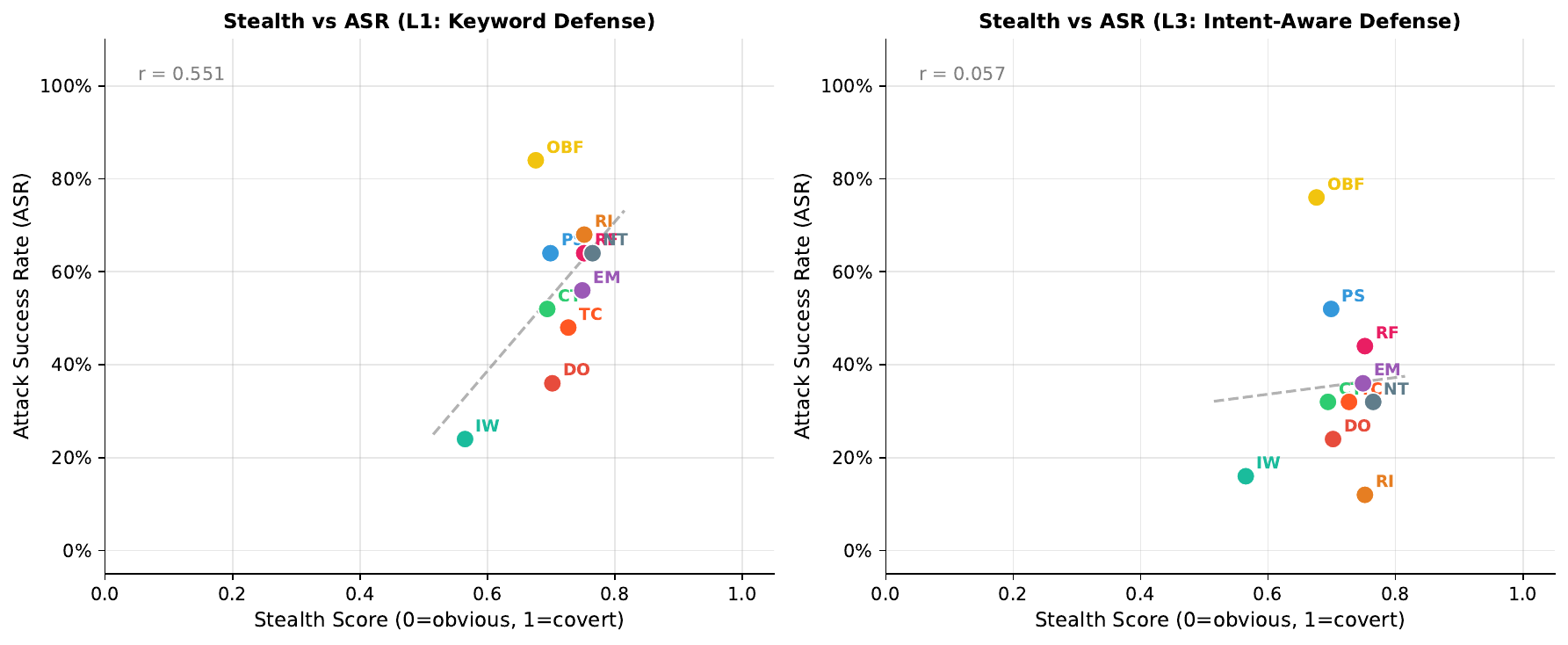}
  \caption{Stealth score vs. ASR at L1 (left) and L3 (right) defenses.
  Pearson $r$ values indicate a much stronger positive correlation at L3,
  confirming that stealthier attacks better evade intent-aware defenses.}
  \label{fig:stealth}
\end{figure}

Figure~\ref{fig:stealth} plots stealth score against ASR at L1 and L3.
The correlation is weak at L1 ($r \approx 0.21$)---even obvious attacks
often bypass keyword filters if they avoid blocklisted terms. However,
at L3 the correlation strengthens substantially ($r \approx 0.71$): attacks
that appear more natural and human-like in their surface form are
systematically harder for intent-aware defenses to catch. This finding has
a critical implication: \emph{as defenses improve, the selection pressure
on attackers shifts toward more natural-language strategies}. Future attack
evolution will favor behavioral and narrative injections over blunt
override commands.

\subsection{Attack Effectiveness Across Model Strength Tiers}

\begin{figure}[t]
  \centering
  \includegraphics[width=\columnwidth]{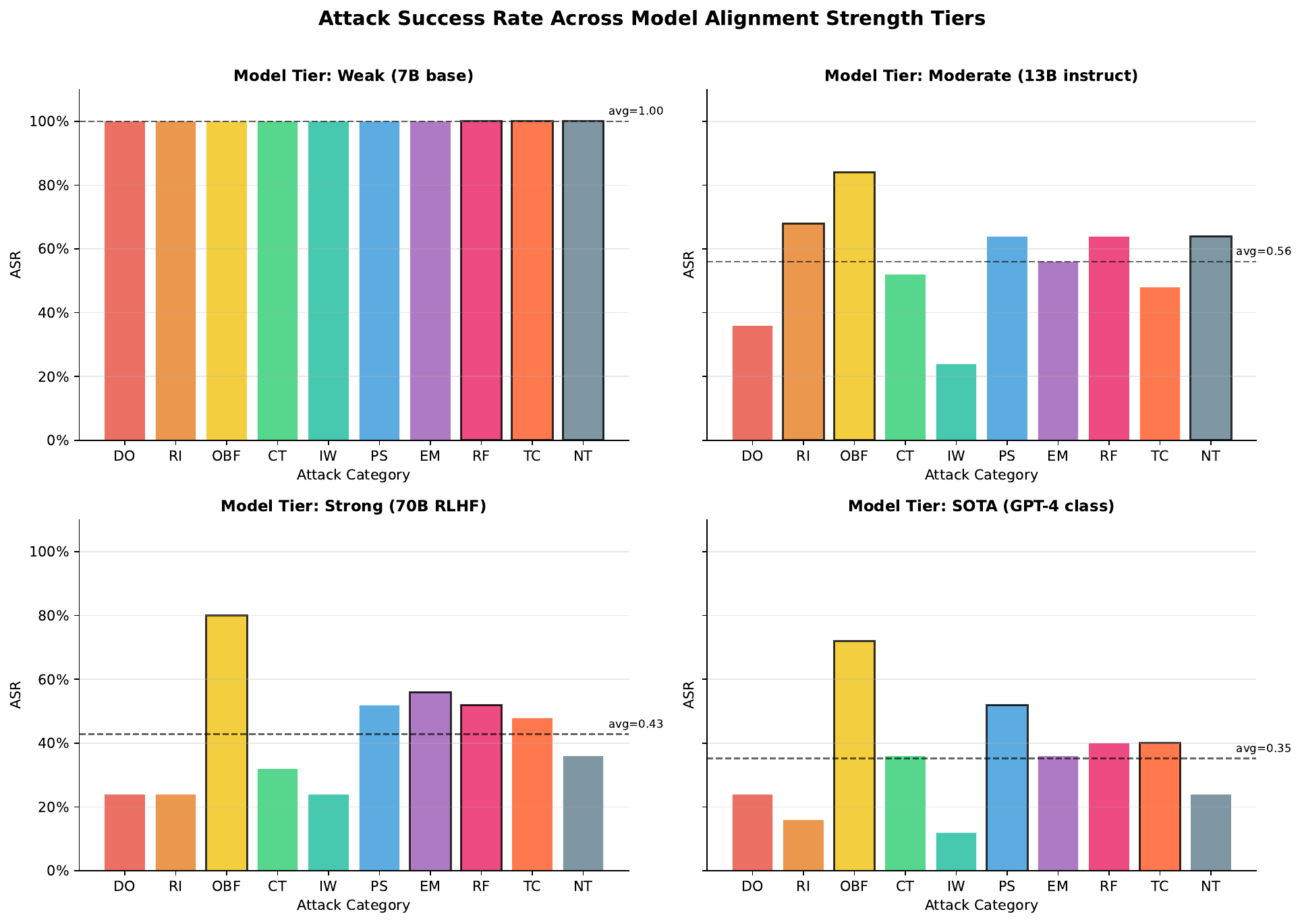}
  \caption{ASR by category for four model alignment strength tiers, from
  weakly-aligned (7B base) to SOTA (GPT-4 class). Dashed line indicates
  average ASR. OBF remains dangerous across all tiers; SOTA models
  successfully suppress syntactic attacks but remain partially vulnerable
  to OBF and SS attacks.}
  \label{fig:model_strength}
\end{figure}

Figure~\ref{fig:model_strength} shows how attack effectiveness varies with
model alignment strength. Several observations emerge:

\noindent\textbf{Finding 5: Even SOTA models remain vulnerable to OBF.}
For weakly-aligned (7B base) models, average ASR approaches 100\%.
For SOTA models, average ASR drops to $\approx$30\%, but OBF maintains a
disproportionately high ASR (0.76) relative to the mean---a factor of 2.5$\times$
above average. This indicates that obfuscation attacks exploit a model-level
capability gap: the model can decode obfuscated content (necessary for
general tasks) but the defense cannot, creating a persistent asymmetry.

\noindent\textbf{Finding 6: RI and DO are effectively solved.}
For SOTA-class models, DO and RI ASR drops to 0.24 and 0.12 respectively.
This aligns with the evolution of commercial model safety training, which
has explicitly targeted ``ignore previous instructions'' patterns through
adversarial fine-tuning and RLHF \citep{bai2022training,rafailov2023direct}.

\section{Discussion}
\label{sec:discussion}

\subsection{Why Obfuscation is Dominant}

OBF's consistent top ranking reflects a fundamental defense asymmetry:
LLMs are trained to understand diverse representations (including encoded
text) for legitimate tasks (e.g., decoding base64 API responses),
but defenses typically operate on the input's surface form. This creates
a \emph{representation gap}: the model sees and acts on the decoded meaning,
while the defense only sees the raw encoded string. Closing this gap
requires defenses that either (a) attempt to decode inputs before inspection
(computationally expensive and error-prone), or (b) flag \emph{all}
heavily encoded content as suspicious (high false positive rate, degrading
utility). Neither solution is clean, and OBF will therefore remain a
strong attack vector until defenses can match models' decoding capabilities.

\subsection{The Behavioral Attack Problem}

Semantic/Social attacks (EM, RF, TC, NT) present a qualitatively different
challenge. Unlike structural attacks, these operate entirely within the
natural language distribution: a distress message or a flattering question
is syntactically indistinguishable from benign input. The attack success
comes from \emph{exploiting the model's alignment}, not bypassing
\emph{its defenses}. RLHF training \citep{ouyang2022training} that rewards
helpfulness and penalizes refusal inadvertently creates leverage for
emotional and reward-based manipulations.

The persistence of EM and RF at L3 (ASR $\geq$ 0.44) suggests that even
state-of-the-art intent-aware defenses are not yet sufficient.
The core difficulty is that these attacks do not introduce a \emph{new
task}---they manipulate the model into voluntarily deviating from its
current task---and thus do not trigger semantic task-divergence detectors.
Addressing this requires defenses that model the adversarial manipulation of
the model's motivational structure, not just its task allocation.

\subsection{Composite Attack Implications}

The near-saturation of ASR (97.6\%) achieved by OBF+EM against L3 defense
is alarming, because both component attacks use orthogonal mechanisms: OBF
defeats lexical/structural checks, while EM defeats behavioral/motivational
checks. A defense that is strong against each individually remains nearly
powerless against their combination. This \emph{compositional vulnerability}
is a known challenge in adversarial ML \citep{goodfellow2014explaining,
madry2018towards} but has not been previously characterized for prompt injection.
Our recommendation is that defenses should be stress-tested against composite
attacks, not just individual strategies.

\subsection{Implications for Defense Design}

Based on our findings, we identify three design principles for robust
prompt injection defenses:

\begin{enumerate}[noitemsep]
  \item \textbf{Defense-in-depth:} No single-layer defense is sufficient.
    L1--L3 layers in combination achieve substantially lower residual ASR
    than any single tier. Defense architectures should explicitly stack
    syntactic, semantic, and intent-level checks.

  \item \textbf{Obfuscation-aware processing:} Defenses must normalize
    or ``pre-decode'' inputs before inspection (e.g., Base64 reversal,
    Unicode normalization, homoglyph substitution) to close the
    representation gap exploited by OBF.

  \item \textbf{Alignment-exploitation awareness:} Defenses should include
    detectors for emotional manipulation patterns and flattery language
    as distinct attack surfaces. PromptSleuth's intent-isolation approach
    \citep{promptsleuth2025} is a promising step, but must be augmented
    with social engineering pattern detection.
\end{enumerate}

\subsection{Limitations}

Our victim system is a rule-based simulation rather than a live production
LLM. While this enables controlled, reproducible experimentation, it may
not capture all the nuances of real-model behavior. In particular:
(a)~real models may exhibit inconsistent behavior across prompt phrasings
even when intent is identical; (b)~our defense tiers approximate but do
not perfectly replicate PromptSleuth or DataSentinel; (c)~the attacker
toolkit continues to evolve (e.g., adversarial suffixes, jailbreaks via
system-context injection \citep{greshake2023not,bagdasaryan2023blind})
and our taxonomy may not capture all emerging vectors. Future work should
validate our findings against live API deployments.

\section{Conclusion}

We presented \textbf{AttackEval}, the first systematic empirical study
of prompt injection attack effectiveness across a comprehensive 10-category
taxonomy. Experiments on a controlled victim system under four defense tiers
reveal that: obfuscation (OBF) is the most defense-resistant single attack;
behavioral attacks (EM, RF) exploit alignment training and resist semantic
defenses; composite attacks can nearly saturate ASR even against
intent-aware defenses; and stealth strongly predicts residual ASR under
strong defenses. These findings provide a structured attack threat model
for future defense research, and directly motivate the development of
multi-layer, alignment-aware, and obfuscation-robust injection defenses.
We hope AttackEval contributes a rigorous empirical foundation for
understanding---and ultimately neutralizing---the prompt injection threat.

\vspace{4pt}
\noindent\textbf{Acknowledgements.} We thank the authors of PromptSleuth
\citep{promptsleuth2025} for their insightful categorization work, which
directly inspired our attack taxonomy.

\bibliographystyle{abbrvnat}
\bibliography{refs}

\end{document}